\documentclass{article}

\usepackage{multirow}
\usepackage{cite}
\usepackage{amsmath,amssymb,amsfonts}
\usepackage{algorithm,algorithmic}
\usepackage{graphicx}
\usepackage{subfigure}
\usepackage{textcomp}
\usepackage[dvipsnames]{xcolor}
\usepackage{comment}
\usepackage{multirow}
\usepackage{booktabs}
\usepackage{array}
\usepackage{algorithm, algorithmic}
\usepackage{pgf}
\usepackage{hyperref}
\usepackage{makecell}
\usepackage{color}
\usepackage{fancyhdr}

\definecolor{gree}{rgb}{0.0,0.7,0}

\title{PhotIQA: A photoacoustic image data set with image quality ratings}
\author{Anna Breger $^{1,2,*}$ \and Janek Gröhl $^{3,4,5}$
    \and Clemens Karner $^{2}$ \and Thomas R Else $^{3,4}$ \and Ian Selby$^{6,7}$  \and {Tom Rix$^{8,9}$} \and {Lara-Sophie Witt$^{5}$} \and {Merle Duchêne$^{5}$} \and Jonathan Weir-McCall$^{6,10}$ \and Carola-Bibiane Schönlieb$^{1}$ }

\date{\footnotesize $^1$University of Cambridge, DAMTP, Cambridge, United Kingdom \\ $^2$Medical University of Vienna, CMPBE, Vienna, Austria \\ $^3$University of Cambridge, Department of Physics, Cambridge, UK \\ $^4$Cancer Research UK Cambridge Institute, Cambridge, UK \\ $^5$ENI-G, a Joint Initiative of the University Medical Center Göttingen and the Max Planck Institute for Multidisciplinary Sciences, G\"ottingen, Germany \\
$^6$University of Cambridge, Department of Radiology, Cambridge, UK \\ $^7$Cambridge University Hospitals, Department of Radiology, Cambridge, UK \\ 
$^{8}$ German Cancer Research Center (DKFZ), IMSY, Heidelberg, Germany\\
$^{9}$ Faculty of Mathematics and Computer Science, Heidelberg University, Germany\\
$^{10}$ King's College London, School of Biomedical Engineering \& Imaging Sciences, London, UK \\
\textsf{$^*$Corresponding author. E-mail: ab2864@cam.ac.uk }}

\begin{document}

\maketitle

\section*{Abstract}
Image quality assessment (IQA) is crucial in the evaluation stage of novel algorithms operating on images, including traditional and machine learning based methods. Due to the lack of available quality-rated medical images, most commonly used full-reference IQA measures have been developed and tested for natural images. Reported pitfalls and inconsistencies arising when applying such measures for medical images are not surprising, as they rely on different properties than natural images. In photoacoustic imaging (PAI), especially, standard benchmarking approaches for assessing the quality of image reconstructions are lacking. PAI is a multi-physics imaging modality, in which two inverse problems have to be solved, which makes the application of IQA measures uniquely challenging due to both, acoustic and optical, artifacts. To support the development and testing of IQA measures we assembled \textbf{PhotIQA}, a data set consisting of 1134 photoacoustic images. The images were rated by {five} experts across five quality properties in a full-reference setting, where the detailed rating enables usage beyond PAI. The data set with the images and corresponding ratings is publicly available on Zenodo. 

\section*{Background}
Advances in medical imaging technologies have been groundbreaking in the last decades, including the rapid development of deep learning methods that operate on a huge amount of image data. To ensure the feasibility of a novel methodology, quantitative image quality assessment (IQA) plays an important role for quality assurance in addition to visual inspection by experts, which is often limited due to time constraints, and therefore IQA measures may even serve as the main assessment criterion. Quantitative IQA measures may be distinguished based on the required information for the assessment step \cite{ssim, perfcomp}. In full-reference (FR) IQA, a reference image is used to evaluate the quality of a corresponding (often degraded) image in a comparative way, relying on a meaningful notion of distance between the two images. In comparison, {no-reference} (NR) IQA aims to judge the quality without a reference based on predefined properties.

Many commonly used IQA measures have been developed for natural images and tested for specific tasks on a small number of publicly available, manually rated data sets. It is unknown how well these measures expand to medical images since they hold very distinct properties, and, moreover, often a different target space (color versus grayscale). Recent research has led to first insights on the applicability of common FR IQA measures to medical imaging data, see e.g.~\cite{10040654, breger2024study21}, showing incompatibilities of popular IQA measures to medical tasks. The research field suffers from the lack of publicly available image data sets with expert annotations, as well as unavailable implementations of published IQA measures. This is reflected in studies that show limitations in the design, e.g., relying on non-expert ratings (see e.g.~\cite{mrinonexpert}), non-realistic distortions (such as Gaussian additive noise, see e.g.~\cite{gaussct}) or a very limited choice of IQA measures (see e.g.~\cite{ssimcompmri}).  Combined with the scarce time resources of medical experts, there are many obstacles for the design of reproducible IQA comparison studies. 

A first step to overcome these constraints is the design and sharing of realistic medical image data sets with expert quality ratings. In the NR setting, recently, there have been first data sets published, see e.g.~a low-dose computed tomography data set~\cite{LEE2025103343}. The data set PhotIQA~\cite{dataPA} is, to the best of our knowledge, the first publicly available, open access data set with medical images rated by experts regarding quality in the FR setting, and specifically, for PAI. Note that due to the nature of FR IQA, the quality ratings may also be employed to assess NR IQA measures when discarding the reference images. 

\subsection*{Photoacoustic imaging and IQA}
Photoacoustic imaging (PAI) is an emerging medical imaging modality with important clinical applications, such as inflammatory bowel disease,  cardiovascular diseases, and breast cancer~\cite{ASSI2023100539}. It is a multi-physics modality, combining interactions of both light and sound with tissue. To reconstruct an image of the parameter of interest in PAI  - the optical absorption coefficient - two inverse problems have to be solved: the acoustic inverse problem to reconstruct an image of absorbed energy density, and the optical inverse problem of recovering optical absorption, scattering, and the Gr\"uneisen parameter. Through limitations in the measurement hardware and non-uniqueness of absorption and scattering, both inverse problems are not trivial to solve, and PAI thus suffers a variety of artifacts that can significantly affect the interpretability and clinical utility of the images~\cite{rietberg2025artifacts,else2025confounding}.

The inverse problems of PAI pertain to accurately visualizing molecular distributions and determining functional tissue information from PA time series measurements~\cite{Janek5}.
It is important to identify suitable measures that can objectively assess the quality of PA image reconstructions, as finding the optimal reconstruction algorithm for the acoustic inverse problem given a certain hardware configuration is crucial for all further steps. 

The state of the art is the employment of common quality measures, such as SSIM \cite{1284395} or PSNR~\cite{grohl2021deep}, which have on the other hand been shown { to not act accurately for many medical image cases}~\cite{Breger:2025vp}. Especially in PAI, with multiple possible device configurations and diverse clinical applications~\cite{parkClinicalTranslationPhotoacoustic2025}, targeted IQA measures must be developed that can accurately assess the image quality given the diagnostic parameter to be extracted from the images. The quality-rated data we present in this paper can be an important first step for initiating experiments that validate the suitability of employed IQA measures. Moreover, the provided detailed quality properties allow a comprehensive insight into task-dependent suitability. As these properties are transferable to other medical imaging tasks, the ratings can also be used for assessing the suitability of IQ measures beyond PA images. 

\section*{Methods}
In this section, we will provide an overview of \textbf{PhotIQA}, the employed image data, and the newly obtained quality ratings.

\subsection*{Photoacoustic Image Data}
\begin{figure}[h!tb]
\centering 
\includegraphics[width=0.23\textwidth]{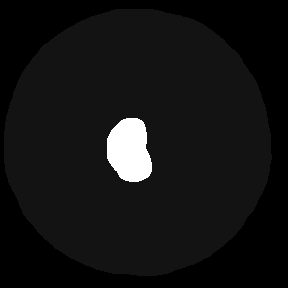} \ 
\includegraphics[width=0.23\textwidth]{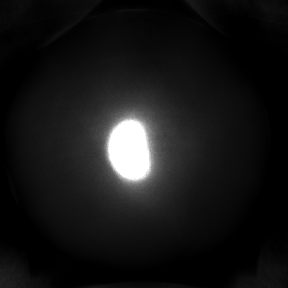} \ 
\includegraphics[width=0.23\textwidth]{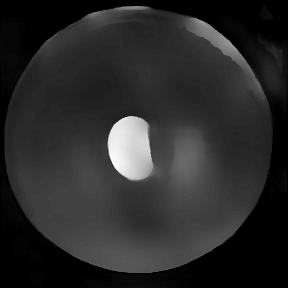} \ 
\includegraphics[width=0.23\textwidth]{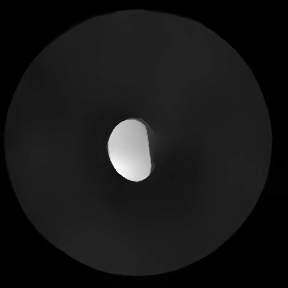} \\
\subfigure[Reference]{\includegraphics[width=0.23\textwidth]{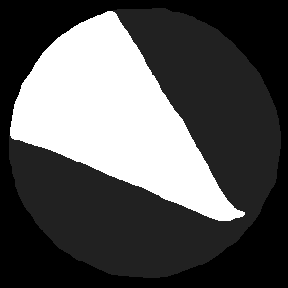}}
\subfigure[\textit{Algorithm 1}]{
\includegraphics[width=0.23\textwidth]{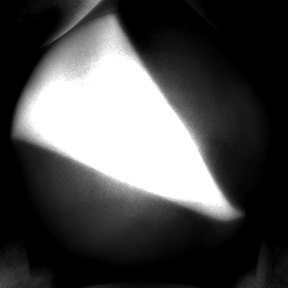}}
\subfigure[\textit{Algorithm 2}]{
\includegraphics[width=0.23\textwidth]{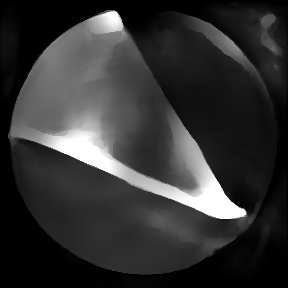}}
\subfigure[\textit{Algorithm 3}]{
\includegraphics[width=0.23\textwidth]{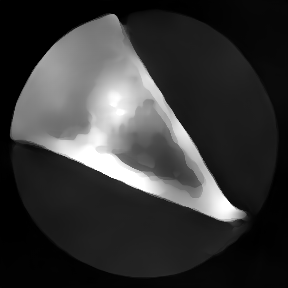}}

\caption{\small Two examples of images in PhotIQA, references (a) and the reconstructions from the described algorithms (b-d). \textit{Algorithm 1} (b) corrects a reconstructed PA image by using the light fluence obtained from simulations. \textit{Algorithm 2} and \textit{Algorithm 3} (c-d) are deep-learning models trained to estimate the absorption coefficient. }
    \label{photoiqa}
\end{figure}

To obtain PhotIQA, we employed a previously published open access data set, cf.~\cite{janekdata}, available via~\cite{grohl_dataset_2023}, that consists of reconstructed PA images containing estimated distributions of the optical absorption coefficient from cross-sectional PA images of piecewise constant test objects (phantoms). The PA data were acquired with a preclinical commercial PA imaging system (MSOT InVision 256-TF, iThera Medical GmbH, Munich, Germany). 
\\
It contains $378$ reference images that have been obtained using a double-integrating sphere \cite{Janek8} setup as a complementary measurement system, which yields point estimates for homogeneous material samples. Because of the piecewise-constant nature of the used phantoms, one can fabricate an additional batch of the material used for the test object, measure it, and relate the calculated properties to the test object. This process is unfeasible for complicated objects or in vivo images, but can serve in this setting to obtain reference images. \\
By applying $3$ different reconstruction methods to the time series data yielding the spatial absorption coefficient images (see Figure~\ref{photoiqa}), we obtain $1134$ reconstructed images, each corresponding to a reference image. The reconstruction method is a two-stage approach, approximating the acoustic inverse problem with a filtered back-projection and then applying one of three methods for the optical inverse problem. These methods were recently published~\cite{janekdata} and can be summarized as follows:  \textit{Algorithm 1} is a fluence compensation algorithm, where the reconstructed image is corrected by a Monte Carlo simulation of the light fluence. The reference absorption and scattering values are obtained through double integrating sphere reference measurements, and this approach is only feasible here through the piecewise-constant nature of the phantoms. \textit{Algorithm 2} is a deep learning algorithm trained in a supervised fashion on purely simulated data, and \textit{Algorithm 3} is architecturally identical to the latter, but trained on experimentally acquired data. The deep learning architecture used for this case was a U-Net, modified for a regression task. The algorithms are described in detail in the publication by Gr\"ohl et al.~\cite{janekdata}.

\subsection*{Image Quality Ratings}
\begin{figure}
\centering
\includegraphics[width=0.98\textwidth]{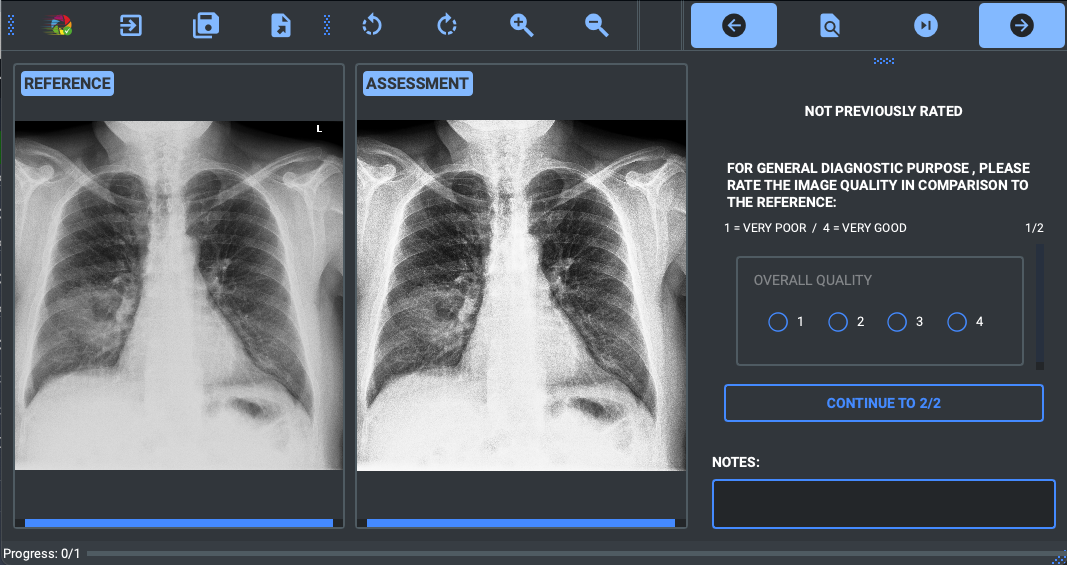}
\caption{The speedyIQA annotation app allows setting a task and rating categories for manual image quality ratings.}
    \label{speedyiqa}
\end{figure}

\begin{figure}[h!tb]
\centering 
\includegraphics[width=0.23\textwidth]{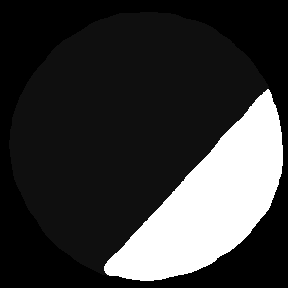}
\hfill 
\includegraphics[width=0.23\textwidth]{Bilder/image366.png}
\hfill
\includegraphics[width=0.23\textwidth]{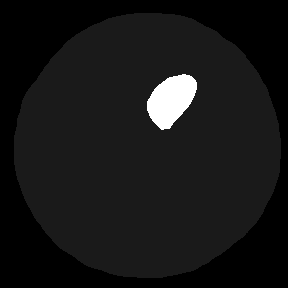} 
\hfill
\includegraphics[width=0.23\textwidth]{Bilder/image261.png} 
\\
\vspace{2mm}

\includegraphics[width=0.23\textwidth]{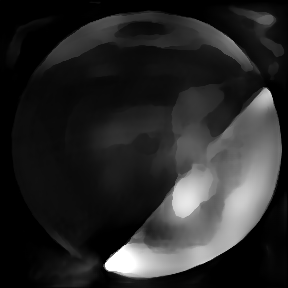}
\hfill
\includegraphics[width=0.23\textwidth]{Bilder/image366_0.png}
\hfill
\includegraphics[width=0.23\textwidth]{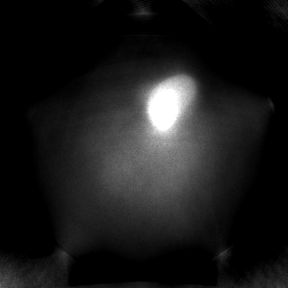} 
\hfill
\includegraphics[width=0.23\textwidth]{Bilder/image261_0.png} 
\\
\vspace{2mm}
\resizebox{0.23\linewidth}{!}{%
\begin{tabular}{lr}
Overall Quality:& 1$|$2\\
Edges: &2$|$2\\
Homogeneity: &1$|$2\\
Intensity$_I$: &1$|$3\\
Intensity$_{BG}$: &2$|$3\\
\end{tabular}
 }
\hfill
 \resizebox{0.23\linewidth}{!}{%
\begin{tabular}{lr}
Overall Quality: &3$|$4\\
Edges:  &3$|$4\\ 
Homogeneity: & 4$|$4\\
Intensity$_I$:& 3$|$4\\
Intensity$_{BG}$:& 3$|$4
\end{tabular}
 }
\hfill
 \resizebox{0.23\linewidth}{!}{%
\begin{tabular}{lr}
Overall Quality: &1$|$2\\
Edges: &1$|$2\\ 
Homogeneity & 1$|$3\\
Intensity$_I$& 2$|$3\\
Intensity$_{BG}$& 1$|$2
\end{tabular}
 }
 \hfill
 \resizebox{0.23\linewidth}{!}{%
\begin{tabular}{lr}
Overall Quality:& $2|4$\\
Edges: &$3|4$\\
Homogeneity: &$1|2$\\
Intensity$_I$: &$2|4$\\
Intensity$_{BG}$: &$3|4$\\
\end{tabular}
 }
\caption{\small {Four examples of the minimum and maximum quality ratings over all annotators for the different detailed quality properties. (Top) reference image and (bottom) reconstructed, assessed image. }}
\label{exlabel}
\end{figure}

{ Five }experts, { with backgrounds in photoacoustic imaging, medical computer science, and medical physics}  have been consulted to rate the reconstructed image data described above consulting a Likert scale, in particular, choosing $1$ (very poor), $2$ (poor), $3$ (good) or $4$ (very good) to describe the quality in comparison to the reference image. The rating was done by { all} experts on $2$ days, taking approximately $9$ hours { per annotator}. They were collected with the publicly available speedyIQA annotation app \cite{speedyiqa}. The software asks the user to set a task and rating categories, see Figure \ref{speedyiqa}. The experts were told not to change the contrast or luminance on their screen during the task. Obtained ratings have been saved in a CSV file, and to account for different rating behaviors between multiple graders, additionally, the z-score has been computed, cf.~\cite{1709988}. Eventually, the mean opinion scores (MOS) between the experts` ratings were computed. In the context of PhotIQA, the experts were asked to rate the images regarding
{
\begin{itemize}
\item Overall Quality, 
\item Edge Visibility (referred to as Edges),
\item Object Homogeneity (referred to as Homogeneity),
\item Object Intensity - Inclusion Area (referred to as Intensity$_I$), and
\item Object Intensity - Background (referred to as Intensity$_{BG}$).
\end{itemize}
}
Here, the object intensity refers to any signals arising from inside the circular test object. Inclusion areas are those areas with increased contrast, and the background refers to the base material that makes up the majority of the test object. See Figure \ref{exlabel} for an example of rated images according to the chosen properties. 

\section*{Data Records}
The PA images ($378$ references and $1134$ corresponding  reconstructions) and the image quality ratings are publicly available at (\url{https://doi.org/10.5281/zenodo.13325196}). The original PA image data is available via the University of Cambridge Apollo repository at (\url{https://doi.org/10.17863/CAM.96644}) and was published by Gr\"ohl et al.~\cite{janekdata} in 2023.

The PA images are stored as 8-bit \textit{png} files of size $288\times288$ pixels in the \textit{zip} file \textit{"normalised\_images.zip"}. {The image normalisation was done by rescaling all pixel values linearly such that the minimum is mapped to 0 and the maximum to 255. The minimum and maximum were defined based on the reference image, i.e.~, all values that were produced by an algorithm and exceeded the maximum of the reference were mapped to 255. This was done to streamline the use of all IQA measures while maintaining quantitative comparability. }This \textit{zip} file contains the folders \textit{"./reference/"}, all reference images named \textit{"image"+\#IMAGE+".png"}, and the folder \textit{"./algorithms/"}, three distorted images for each reference image named \textit{"image"+\#IMAGE+"\_"+\#ALGORITHM+  }\textit{".png"} corresponding to the three employed algorithms described above.

{
The ratings of each quality property are stored in the file \textit{annotations.zip}, which contains \textit{csv} files named \textit{"annotations\_"+TASK}. These files contain the columns 
\begin{itemize}
    \item "filename": File name of the distorted image file
    \item "rater1" ,..., "rater5": Ratings from the 5 experts
    \item "mos": MOS of all 5 ratings
    \item "z-score rater1" ,..., "z-score rater5": Z-scores of all 5 ratings
    \item "z-score mos": MOS of the z-scores of all 5 ratings
\end{itemize} 
}
{
The file \textit{"IQA\_evaluation.zip"} contains the script \textit{"main.ipynb"}, which allows to directly reproduce results of Table \ref{IQAeva}, as well as the text file \textit{"software\_versions.txt"}, a list of the versions of all employed IQA metrics.}

\section*{Technical Validation}\label{techval}
\subsection*{Inter-rater consistency}
\begin{figure}[h!tb]
\centering 
\includegraphics[width=0.7\textwidth]{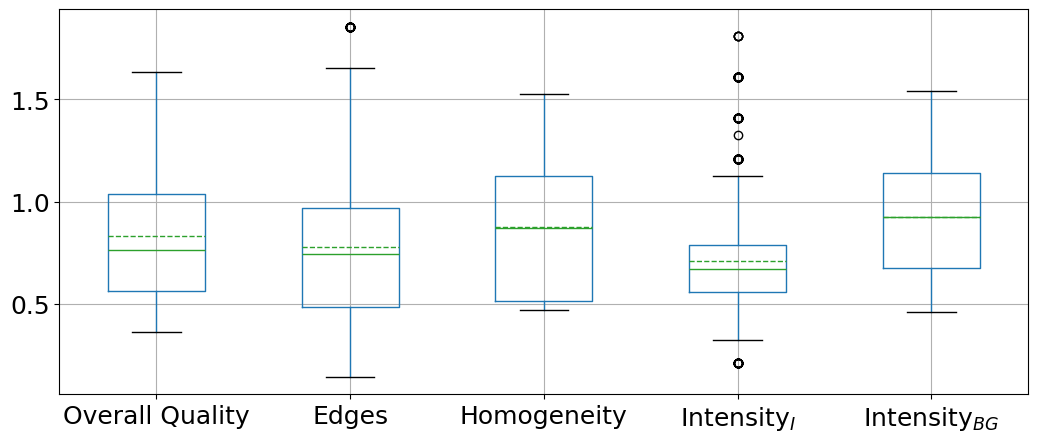}
\includegraphics[width=0.7\textwidth]{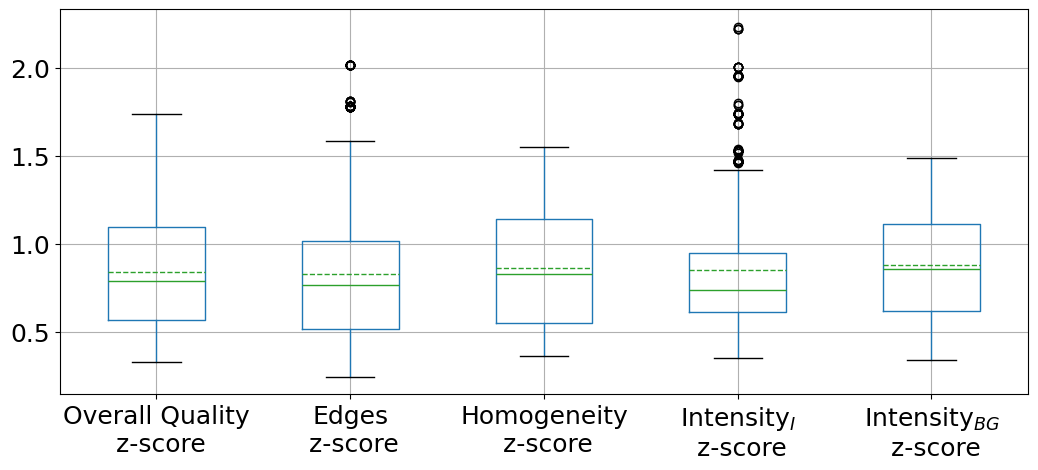}
\caption{\small {Box plot of the mean absolute differences (top) and the absolute differences of z-scores (bottom) of all $5$ raters to the (z-scored) MOS with the median (green line) and mean (striped green line).}}
\label{boxp}
\end{figure}

\begin{figure}[h!tb]
\centering
\includegraphics[width=0.23\textwidth]{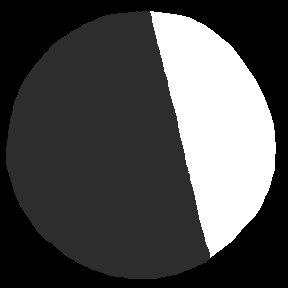} 
\hfill
\includegraphics[width=0.23\textwidth]{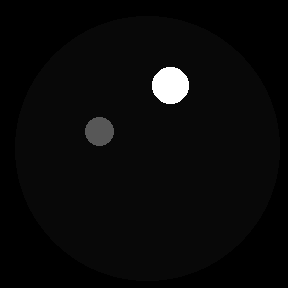} 
\hfill
\includegraphics[width=0.23\textwidth]{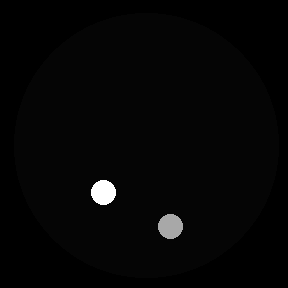}
\hfill
\includegraphics[width=0.23\textwidth]{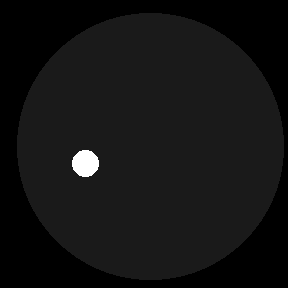}
\\
\vspace{2mm}
\includegraphics[width=0.23\textwidth]{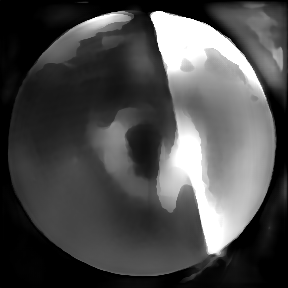} 
\hfill
\includegraphics[width=0.23\textwidth]{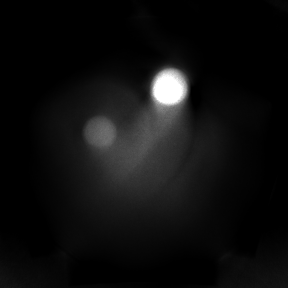} 
\hfill
\includegraphics[width=0.23\textwidth]{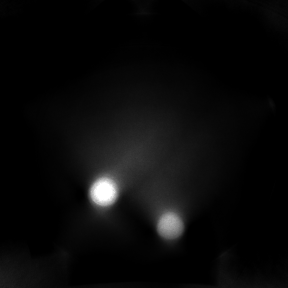}
\hfill
\includegraphics[width=0.23\textwidth]{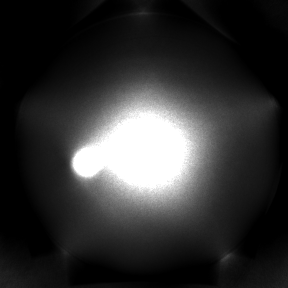} 
\\
\vspace{2mm}
\resizebox{0.23\linewidth}{!}{%
\begin{tabular}{lr}
Overall Quality: &1$|$2\\
\textbf{Edges:} &1$|$4\\ 
Homogeneity & 1$|$2\\
Intensity$_I$&2$|$3\\
Intensity$_{BG}$&1$|$2
\end{tabular}
}
\hfill
 \resizebox{0.23\linewidth}{!}{%
\begin{tabular}{lr}
Overall Quality: &1$|$3\\
\textbf{Edges:} &1$|$4\\ 
Homogeneity & 1$|$3\\
Intensity$_I$&3$|$4\\
Intensity$_{BG}$&2$|$2
\end{tabular}
}
\hfill
\resizebox{0.23\linewidth}{!}{%
\begin{tabular}{lr}
Overall Quality:& 2$|$3\\
Edges: &1$|$3\\
Homogeneity: &2$|$3\\
\textbf{Intensity$_I$}: &1$|$4\\
Intensity$_{BG}$: &2$|$3\\
\end{tabular}
 }
\hfill
 \resizebox{0.23\linewidth}{!}{%
\begin{tabular}{lr}
Overall Quality: &1$|$1\\
Edges:  &1$|$1\\ 
Homogeneity: & 1$|$2\\
\textbf{Intensity$_I$}:& 1$|$4\\
Intensity$_{BG}$:& 1$|$1
\end{tabular}
 }
\caption{\small { Four examples of the minimum and maximum quality ratings from all annotators corresponding to box plot outliers in Figure \ref{boxp} (properties in bold). Reference image (top) and reconstructed, assessed image (bottom).} }
\label{boxp_examples}
\end{figure}

 { First, we study the IQA ratings provided by the experts. In Table \ref{corrtab} and Figure \ref{boxp} we verify the agreement of ratings among them. The results in the table demonstrate a high correlation between all raters and tasks, and the latter shows the box plot of the absolute differences between the raters, for the raw data as well as the z-scores. We can see directly that the mean and median absolute difference of the raw ratings is at most $1$, indicating a high alignment between the experts. } The z-scores compensate for biases of individual raters towards low or high scores that may limit the direct interpretability. For example, the meaning of the rating (4) might differ between the experts depending on a tendency towards low or high ratings, and the z-scores allow insights into the corrected difference.
{
In the box plot corresponding to the z-scores, we can observe that the agreement acts similarly across the properties. The categories "Intensity$_{I}$" and "Edges" contain the most outliers. Visual examples of extreme outliers in those categories are provided in Figure \ref{boxp_examples}.
}

\begin{figure}[h!tb]
\centering 
\subfigure[Overall quality]{\includegraphics[width=0.45\textwidth]{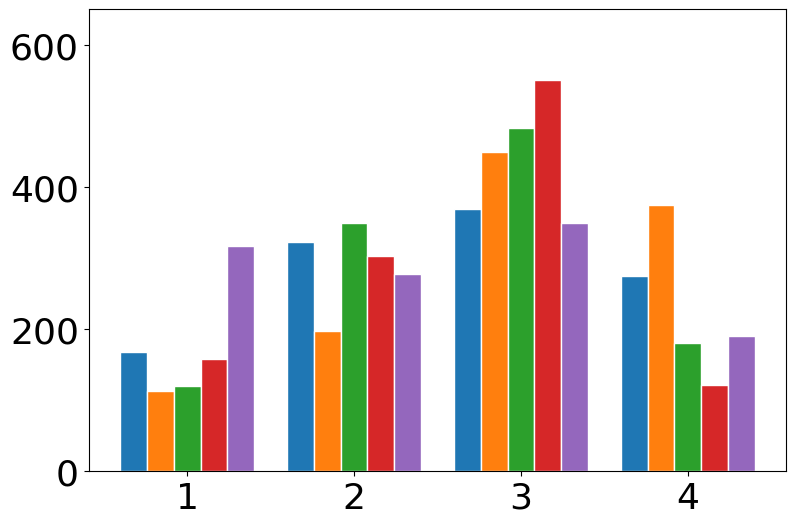}}
\subfigure[Edges]{\includegraphics[width=0.45\textwidth]{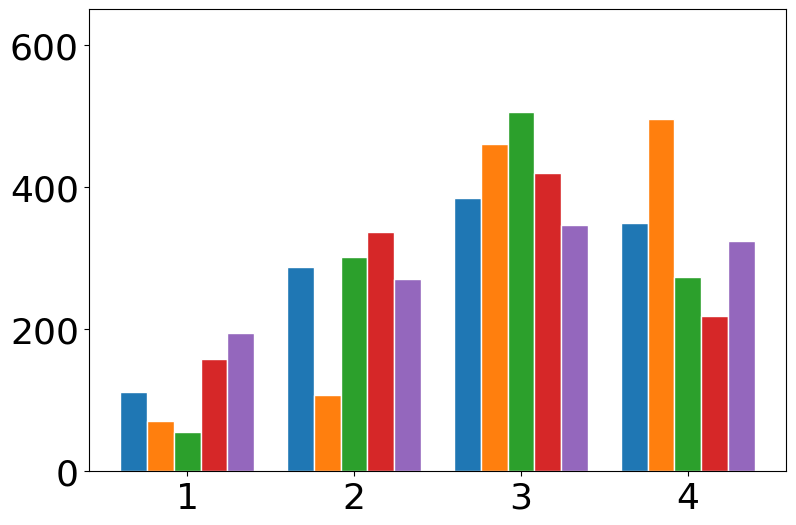}}
\subfigure[Homogeneity]{\includegraphics[width=0.45\textwidth]{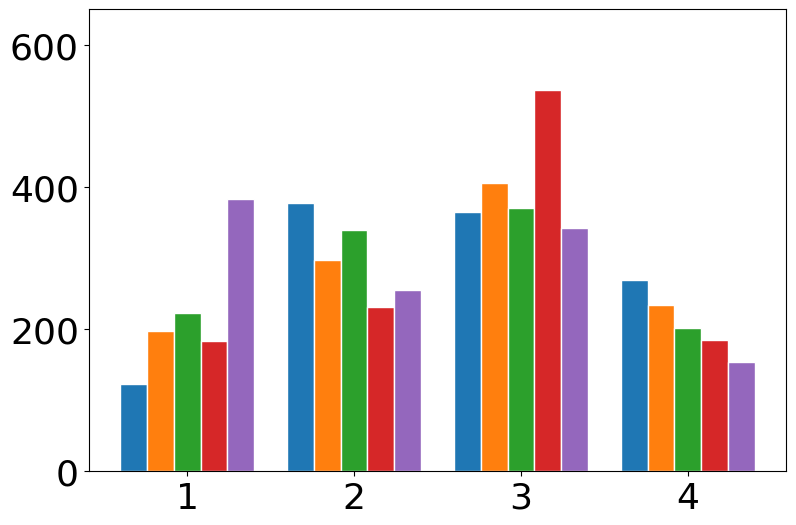}}
\subfigure[Intensity$_{I}$]{\includegraphics[width=0.45\textwidth]{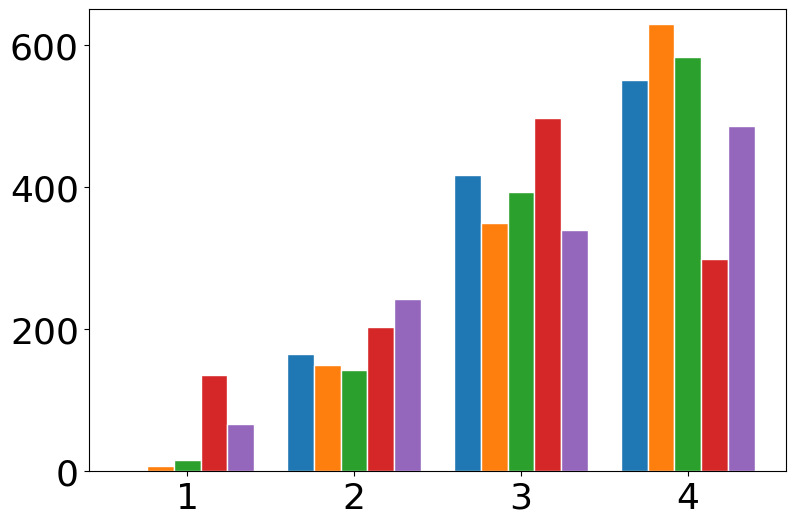}}
\subfigure[Intensity$_{B}$]{\includegraphics[width=0.45\textwidth]{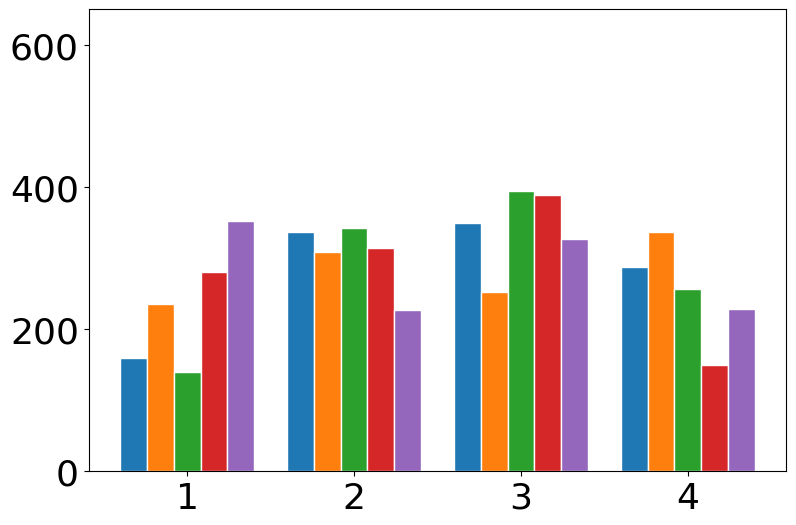}}
\caption{\small {The distribution of the ratings for each quality property (a)-(e) plotted for the 5 annotators.} }
\label{distribution}
\end{figure}

Moreover, in Figure \ref{distribution}, we show the amount of each rating given per task. We observe that the rating (1) was given much less than the ratings (2-4), which is caused by the reconstructed images stemming from three reconstruction algorithms that all give somewhat reasonable results. {The property "Intensity$_I$" peaks at the highest rating (4). %, due to the design of the phantoms. 
The other properties show behavior closer to a normal distribution peaking at the rating (3). Rater $3$ showed a different rating behavior regarding (4), which they used much less than the other raters, and rater $2$ leaned towards higher ratings. The z-score accounts for those personal tendencies. }

\begin{table}[]
    \centering
\begin{tabular}{l|rrrrr|l}
\toprule
 & Rater 1 & Rater 2 & Rater 3 & Rater 4 & Rater 5 & Mean \\
\midrule
Overall Quality & 0.86 & 0.86 & 0.90 & 0.86 & 0.89 & 0.88 \\
Edges & 0.87 & 0.87 & 0.83 & 0.87 & 0.90 & 0.87 \\
Homogeneity & 0.90 & 0.90 & 0.88 & 0.83 & 0.87 & 0.88 \\
Intensity$_{I}$ & 0.86 & 0.83 & 0.79 & 0.84 & 0.85 & 0.83 \\
Intensity$_{B}$ & 0.85 & 0.84 & 0.92 & 0.88 & 0.89 & 0.88 \\
\midrule
Mean & 0.87 & 0.86 & 0.87 & 0.85 & 0.88 &  \\
\bottomrule
\end{tabular}
    \caption{The table displays the Spearman Rank Correlation Coefficient (SRCC) for each task and rater, calculated between each rater's annotations and the z-scored MOS. The last column states the average SRCC value for each task. The last row states the average SRCC value for each rater.}
    \label{corrtab}
\end{table}


\subsection*{IQA evaluation}
{To provide a baseline assessment of the validity of common IQA measures, we compare their values to the manual ratings. The chosen IQA measures were computed whenever possible with implementations provided by their authors using default parameters, either in MATLAB or Python, and in the case for PSNR, SSIM and MS-SSIM, we utilise commonly used Python and Matlab implementations in their default settings. The exact versions and software packages are stated as part of the data set on Zenodo, see Section \textit{Data Records} above for more details. 

For the evaluation of the IQ measures' performance, we employ the Spearman Rank Correlation Coefficient (SRCC) and the Kendall Rank Correlation Coefficient (KRCC) on the whole image dataset, which assess the rank between the IQA measures and the manual ratings. 
The absolute SRCC and KRCC between the MOS of the z-scores and the IQA measures are stated in Table \ref{IQAeva}. 
{We can observe that MS-SSIM \cite{msssim}, with the implementation from the Python torchmetrics library, is among the top 3 for all properties. It is important to notice that there is a relevant difference to the implementation provided in MATLAB. In line with previous experiments, HaarPSI\cite{Reisenhofer18}, especially in the HaarPSI$_{\text{med}}$ \cite{haarpsimed} setting, as well as IW-SSIM \cite{iwssim}, yield also good results in all categories. LPIPS \cite{lpips} and GMSD \cite{gmsd} show promising behavior for some properties. The commonly used measures PSNR and SSIM struggle for almost all quality properties. PSNR is the best measure regarding "Intensity$_I$", which is not surprising, as it directly computes the pixel-wise difference and is sensitive to shifts in high intensity values. 

{ Other FR IQA measures include FSIM \cite{5705575}, MDSI \cite{mdsi} and DISTS \cite{9298952}.} The tested NR measures {NIQE \cite{niqe}, PaQ-2-PiQ \cite{paq2piq}, TeamEpoch \cite{LEE2025103343} and UIQA \cite{UIQA}} yield generally lower results, which is not surprising, as this a more distinct and different task than the direct comparison to a reference image. It has also to be noted that the images were annotated in a FR and not NR setting, positively biasing the evaluation for FR measures.}

\begin{table}
   \resizebox{\linewidth}{!}{%
\begin{tabular}{lccccc}
\toprule
 & Overall Quality & Edges & Homogeneity & Intensity$_{I}$ & Intensity$_{BG}$ \\
 \midrule
PSNR* & 0.79/0.59 & 0.63/0.45 & 0.82/0.62 & \textbf{\color{Emerald}0.74/0.56} & 0.69/0.51 \\
SSIM* & 0.70/0.53 & 0.67/0.51 & 0.65/0.47 & 0.46/0.32 & 0.72/0.53 \\
HaarPSI$_{med}$ & \textbf{\color{Emerald}0.88/0.71} & \textbf{\color{Emerald}0.82/0.63} & 0.83/0.63 & 0.60/0.42 & \textbf{\color{Emerald}0.86/0.66} \\
HaarPSI & \textbf{\color{Emerald}0.87/0.68} & 0.79/0.59 & 0.83/0.62 & 0.63/0.45 & 0.81/0.61 \\
MS-SSIM* & \textbf{\color{Emerald}0.87/0.70} & 0.74/0.56 & \textbf{\color{Emerald}0.86/0.67} & 0.59/0.41 & \textbf{\color{Emerald}0.89/0.72} \\
MS-SSIM & \textbf{\color{Emerald}0.91/0.74} & \textbf{\color{Emerald}0.80/0.61} & \textbf{\color{Emerald}0.89/0.71} & \textbf{\color{Emerald}0.67/0.48} & \textbf{\color{Emerald}0.88/0.71} \\
IW-SSIM* & \textbf{\color{Emerald}0.87/0.69} & 0.77/0.58 & \textbf{\color{Emerald}0.85/0.65} & 0.65/0.46 & \textbf{\color{Emerald}0.86/0.67} \\
GMSD* & 0.83/0.64 & 0.68/0.49 & \textbf{\color{Emerald}0.85/0.67} & \textbf{\color{Emerald}0.68/0.49} & 0.77/0.58 \\
FSIM* & 0.83/0.65 & 0.70/0.52 & 0.84/0.64 & 0.64/0.45 & 0.82/0.62 \\
MDSI* & 0.75/0.55 & 0.65/0.47 & 0.74/0.54 & 0.66/0.47 & 0.65/0.46 \\
LPIPS$_{Alex}$ & 0.82/0.63 & \textbf{\color{Emerald}0.81/0.63} & 0.74/0.55 & 0.47/0.32 & \textbf{\color{Emerald}0.86/0.67} \\
DISTS* & 0.76/0.59 & 0.67/0.50 & 0.75/0.57 & 0.45/0.32 & 0.82/0.64 \\
\midrule
NIQE* & 0.56/0.39 & 0.34/0.23 & 0.70/0.51 & 0.53/0.38 & 0.54/0.38 \\
PaQ-2-PiQ & 0.30/0.20 & 0.51/0.35 & 0.12/0.08 & 0.05/0.04 & 0.33/0.22 \\
TeamEpoch & 0.25/0.17 & 0.39/0.27 & 0.09/0.06 & 0.04/0.02 & 0.33/0.22 \\
UIQA & 0.52/0.35 & 0.58/0.40 & 0.43/0.28 & 0.25/0.17 & 0.48/0.33 \\
\bottomrule
\end{tabular}
}
\caption{Baseline results: SRCC/KRCC between the MOS of the manual expert ratings' z-scores and common full-reference (top) and no-reference (bottom) IQA measures. $^*$ denotes that implementations were provided in MATLAB rather than Python. The $3$ highest results are bold and colored.}
\label{IQAeva}
\end{table}

\section*{Usage notes}
In order to pair the images with the scores of a specific property, e.g.~overall quality, we employ the \textit{csv} file \textit{"annotations\_overall\_quality.csv"}. The filename leads to the corresponding \textit{png} images in the folders \textit{"./algorithms/"} and  \textit{"./references/"} contained in the \textit{zip} file \textit{"normalised\_images.zip"}.

\subsection*{Limitations}
{The ratings were obtained by 5 experts and, as expected for manual annotations, show some variability in rating behavior, see Figure \ref{boxp_examples}. Nevertheless, we also show in Table \ref{corrtab} and Figure \ref{boxp}, that, overall, the ratings correlate highly between all raters.

The two-stage PA reconstruction was done using filtered back-projection for the acoustic inverse problem. While we expect a deep learning based approach for the optical inversion to learn to account for systematic reconstruction artifacts, the concrete algorithm for the acoustic inversion is expected to have an impact~\cite{grohl2025digital}. Moreover, in this work, we evaluate the PAI reconstruction performance purely on the basis of comparing reconstructed absorption coefficients to reference absorption coefficients. In a clinical setting, this might not always be the most important endpoint. Furthermore, the narrow endpoint limits the spectrum of PAI artifacts represented by this dataset. 

{ A major benefit of the dataset is the access to known underlying tissue properties obtained with established reference measurement methods, which allows the benchmarking of FR IQA measures. However, the clinical meaningfulness is limited; on a clinical dataset, one would not have access to such ground-truth knowledge. This hinders the use of FR image quality measures. If the true image reconstruction is unknown, this leaves only NR (or reduced-reference) measures.  Such measures could then be benchmarked against medical expert annotators as well as clinical outcome measures.

We normalise the images between 0 and 255 given the bounds of the reference image. While this streamlines the application of the IQA measures, it simultaneously introduces a bias, where systematic over/underestimations of the algorithms may be masked. Because of this reason, we did not introduce classical quantitative distance measures, such as the root mean squared error or absolute/relative error measures, and suggest computing such measures on the unnormalised images available in the original data publication~\cite{janekdata,grohl_dataset_2023}. }
{
\section*{Data Availability}
The normalised images and manual quality ratings are publicly available under a Creative Commons license (CC-BY 4.0) at \url{https://doi.org/10.5281/zenodo.13325196}~\cite{dataPA}.
The original phantom data is publicly available under a Creative Commons license (CC-BY 4.0) at \url{ https://doi.org/10.17863/CAM.96644}~\cite{grohl_dataset_2023}.\\}
{
\section*{Code Availability}
The code for the PAI reconstruction algorithms is available open source at (\url{https://github.com/BohndiekLab/end_to_end_phantom_QPAT}) under an MIT license. The Python code of the IQA experiments, including the implementations of the Python based IQA measures, is available on GitHub \url{https://github.com/ideal-iqa/iqa-eval} under an MIT license. { A Python file to reproduce results in Table \ref{IQAeva} is available under a Creative Commons license (CC-BY 4.0) at \url{https://doi.org/10.5281/zenodo.13325196}~\cite{dataPA}.} The speedyIQA annotation app is available at \url{https://github.com/selbs/speedy_iqa} under an MIT license.} }

\section*{Acknowledgments}
The authors wish to acknowledge support from the EU/EFPIA Innovative Medicines Initiative 2 Joint Undertaking - DRAGON (101005122) (A.Br., I.S., C.B.S.); the Austrian Science Fund (FWF) through project T1307 (A.Br., C.K.); the German Research Foundation through the grant GR 5824/1 (J.G.) and project number 462569370 (T.R.); EPSRC UK EP/X037770/1 (T.R.E.); Cancer Research UK through C9545/A29580 (T.R.E); the ERC under the European Union's Horizon research and innovation programme through project NEURAL SPICING (grant 101002198) (T.R.),the NIHR Cambridge Biomedical Research Centre (BRC-1215-20014) (I.S.) and (NIHR203312) (C.B.S); C.B.S also acknowledges support from the Philip Leverhulme Prize, the Royal Society Wolfson Fellowship, the EPSRC (EP/V029428/1, EP/V026259/1, EP/S026045/1, EP/T003553/1, EP/N014588/1, EP/T017961/1, the Wellcome Innovator Awards 215733/Z/19/Z and 221633/Z/20/Z, the EPSRC funded ProbAI hub EP/Y028783/1), and the European Union Horizon 2020 research and innovation programme under the Marie Skodowska-Curie grant agreement REMODEL. 

Please note that the content of this publication reflects the authors’ views and that neither NIHR, the Department of Health and Social Care, IMI, the European Union, EFPIA, nor the DRAGON consortium are responsible for any use that may be made of the information contained therein.

{
\section*{Contributions}
A.B., J.G., T.R.E. and I.S. designed the study and performed the data collection; A.B. and J.G. wrote the initial version of the paper; J.G., T.R.E., T.R., L.-S.W., and M.D. performed data annotation; C.K. and A.B. worked on the technical validation and figures. All authors discussed the results and contributed to the editing of the final manuscript.}

\section*{Ethics declaration}
The authors declare no competing interests.

\small
\bibliographystyle{splncs04}
\bibliography{bib1}

\end{document}